# Refined scenario of the collinear cluster tri-partition "Ni-bump" mode

Yu. V. Pyatkov[1,2], D. V. Kamanin[2], Yu. M. Tchuvil'sky[3], A. A. Alexandrov[2], I. A. Alexandrova[2], Z. I. Goryainova[2], R. Korsten[4], V. Malaza[5], E. A. Kuznetsova[2], A. O. Strekalovsky[2], O. V. Strekalovsky[6,2], Sh. Wyngaardt[4], and V. E. Zhuchko[2]

[1]National Nuclear Research University MEPhI (Moscow Engineering Physics Institute), Moscow, Russia
[2]Joint Institute for Nuclear Research, Dubna, Russia
[3]Skobeltsyn Institute of Nuclear Physics, Lomonosov Moscow State University, 119991 Moscow, Russia
[4]University of Stellenbosch, Stellenbosch, Western Cape, South Africa
[5]University of Stellenbosch, Faculty of Military Science, Military Academy, Saldanha 7395, South Africa
[6]Dubna State University, 141980 Dubna, Russia

**Abstract.** In our previous publications, we discussed various manifestations of a new decay channel of the low excited heavy nuclei dubbed collinear cluster tri-partition (CCT). The most populated CCT mode was revealed in the mass correlation distribution of fission fragments as a local region ("bump") of increased yields below the loci linked to the conventional binary fission. The bump was dubbed "Ni-bump" because it is centered at the masses associated with the magic isotopes of Ni. We have essentially modified hardware and software used in our previous experiments. New experimental data obtained with three time's larger statistics confirm the structure peculiarities of the Ni-bump. In this work, we also propose a refined qualitative model of the Ni-bump. The CCT in this mode is supposed to occur as a two stages sequential decay process of the very deformed pre-scission configuration. Coulomb inelastic scattering of the intermediate fragment being in the weakly bound shape isomer state leads to its break-up in the dense medium.

## 1 INTRODUCTION

In our major publications dedicated to collinear cluster tri-partition (CCT) of heavy low excited nuclei [1–3], we have presented multiple experimental indications of this new type of multibody decay. In the most recent paper [4], we have discussed the physical scenario behind the so-called "Ni-bump". This most populated CCT mode revealed so far manifests itself in the mass correlation distribution of fission fragments (FFs) via linear structures $M$ = const, with the constants corresponding to the masses of the magic nuclei of $^{68,72}$Ni and $^{128}$Sn. After or almost simultaneously with our paper [4] several theoretical works dedicated or related to the CCT were published.

In ref. [5] the limits imposed by quantum mechanical principles on angular distribution of a tri-partition process were obtained. The authors came to the conclusion that only the sequential and "almost sequential" mechanisms may result in quasi-collinear events, but in any case, the estimated lower limit of the width of the angular distribution for the light ("Ni-like") CCT partner exceeds experimental value. Nevertheless, the limit obtained in ref. [5] is much less pessimistic when compared to that from ref. [6]. Very specific argument in favor of the CCT channel in $^{238}$U was found in ref. [5] using the geochemical data. Abnormal abundance of $^{38}$Ar in various samples of the uranium ore agreed with that typical for the CCT by the yield.

Ternary fission of $^{252}$Cf was studied in ref. [7] using three-cluster and unified ternary fission models based on classical physics concepts. Collinear and equatorial pre-scission configurations were analyzed in ternary fission with the $^{10}$Li as the third fragment. For the fragment combination showing the highest relative yield, both models predict collinear configuration.

Scission point model was applied in ref. [8] for the mass distribution of ternary fission in $^{235}$U ($n_{th}$, f) reaction. The obtained results indicate that the CCT mode associated with double closed shell $^{132}$Sn is achievable if the process of ternary fission happens in two stages: i) formation of $^{132}$Sn plus elongated partner; ii) subsequent breaking of the elongated partner into another closed shell nucleus $^{48}$Ca and the remaining fragment of $^{56}$Ti.

The trajectory calculations carried out in ref. [9] allowed numerically estimating the kinetic energy and position of different ternary fission fragments, and the set of initial parameters including the initial emission angle of ternary particle. The initial angle was measured with respect to the axis connecting the centers of the side heavy fragments. The authors note: "As the size of the third fragment increases, the orthogonal and/or equatorial type emission is found to vary as the initial angle varies. In particular, for the lowest initial angle of 1° considered for the present study, the trajectory is found to be collinear supporting the experimental claims of collinear cluster tripartition."

True ternary fission of $^{252}$Cf was a subject of critical examination by means of the trajectory calculations in ref. [10]. The authors calculated angular distribution of the fragments with the aim to study the effect of the spins of the fragments arising at the moment of scission. Both scissions were assumed to occur nearly simultaneously, as opposed to our experimental findings. The total spin is compensated by the orbital angular momentum of the relative motion of the fragments. The orbital momentum gives evidence of the initial velocities of the FFs, which are perpendicular to the fission axis. The fragments are supposed to stay collinear on a rotating fission axis. This assumption is not obvious for the ternary system but just the allowance for the initial transverse velocity results in the final angular divergence of the fragments. The predicted angular divergence of the CCT partners is very low and does not exceed one degree at reasonable initial conditions.

For the first time, the study of the excitation energy-dependent potential-energy surfaces for both the spherical and deformed fragments from the ternary fragmentation of $^{252}$Cf at four different excitation energies of the fissioning nuclei was performed in ref. [11]. For the spherical fragments, some energy maxima were obtained around magic and/or semi-magic numbers of nuclei. Calculations for deformed ternary fragments showed the energy maximum around the charge numbers of the fragments $Z_3 = 16$ with $Z_2 = 38 - 34$ and $Z_1 = 44 - 48$ regions due to the presence of higher $\beta_2$ deformation values.

The theoretical descriptions of the decay of heavy nuclei into three comparable fragments proposed so far are summarized in the just published review of W. von Oertzen and A. K. Nasirov [12]. In almost all the publications sited there, the stage that precedes the formation of the ternary nuclear system (TNS) is not studied. The potential energy surface of the TNS consisting of spherical or slightly deformed nuclei is under analysis. For the ternary partitions into fragments of comparable masses collinear chain-like prescission configuration is predicted. The ternary decay of such system is supposed to be sequential i.e. to be a sequence of two two-body decays. The calculations show that the central fragment in the chain attains energies close to zero. Our results obtained in the CCT dedicated experiments are also discussed in the review but the interpretation of the authors is only partially consistent with ours. It seems further efforts required to build a complete self-consistent model of the CCT.

Mass distributions of fragments from *ternary fission of superheavy nuclei* are investigated with the aid of the statistical theory in ref. [13]. This is another work in a whole series of the theoretical publications (see references in work [13]) devoted to the ternary decay channel in super heavies. Our results in studying of the CCT are used for testing the theoretical models developed.

We would like to refer to the work [14] from the related field of physics. The resonance scattering of diatomic molecules (dimers) by atoms via three-particle metastable trimer states and the molecular dissociation induced by collisions with atoms were considered. After potential replacement, the developed approach could be useful for analysis of interaction of the CCT-born di-nuclear systems when they pass through the medium of the source backing. The authors of work [14] also mention the CCT as a field of application of the developed methods.

As emphasized in our publications [1, 2], we have only observed the Ni-bump in the spectrometer arm facing the Cf source backing. So far, this peculiarity was beyond the scope of the theoretical considerations, but it likely plays a decisive role in observing the effect, as will be demonstrated below. Another motivation for this paper is to present our new experimental results, obtained recently.

**2 EXPERIMENTS**

Most of our experiments in recent years were dedicated to searching for the rare decay modes of low excited actinide nuclei while paying special attention to reliability of identification of such fission events. To increase the reliability, the shapes of the signals are digitized with off-line processing of the signal images. Here we present the results of the experiment (Ex1) performed using such data processing approach. The scheme of the setup is shown in fig. 1. It is the double-armed time-of-flight spectrometer of fission fragments (FFs). Two "start" timing detectors on the microchannel plates $St$1 and $St$2 were placed in the center of a vacuum chamber, with a $^{252}$Cf (sf) source on a 50 μg/cm$^2$ thick Al$_2$O$_3$ backing between them. The source was made by stippling. The active spot of the source 5 mm in diameter is located in the center of the backing, facing arm-2 of the spectrometer. In two similar timing detectors (fig. 1(b)), the fragment passes through a 32 µg/cm$^2$ thick Lexan foil with a 40 µg/cm$^2$ thick gold layer on it.

Four mosaics of eight PIN diodes each provided measurement of both the FF energy and time-of-flight. The mean flight-path for each mosaic did not exceed 15 cm and the working area of the PIN diode was 1.4x1.4 cm$^2$. The open surfaces of the adjacent diodes in the mosaic are separated from each other by an opaque for fragments strip of 5 mm width, or 2$^0$ angular strip size.

The data acquisition system consisted of the multichannel fast flash-ADC (Amplitude to Digital Convertor) digitizer CAEN DT574, logic blocks providing trigger signals, and a personal computer. Current value of the signal was measured by the digitizer every 0.2 ns. Time reference point on the PIN diode signal was calculated using new algorithm proposed earlier [15]. The algorithm fits the initial part of a leading edge of the PIN diode signal with parabola function, under condition that the parabola vertex lays on the mean value of the signal's base line. The parameters of the function are calculated using $\chi^2$ minimization of several points at the leading edge above three sigma (standard deviation of the base line values) level. Experimental testing at accelerator beam proved that this method gives unbiased, "true", time reference corresponding to the real start of the signal [15]. Thus, skewness of the time-of-flight due to so called "plasma delay" (PD) [16] is eliminated.

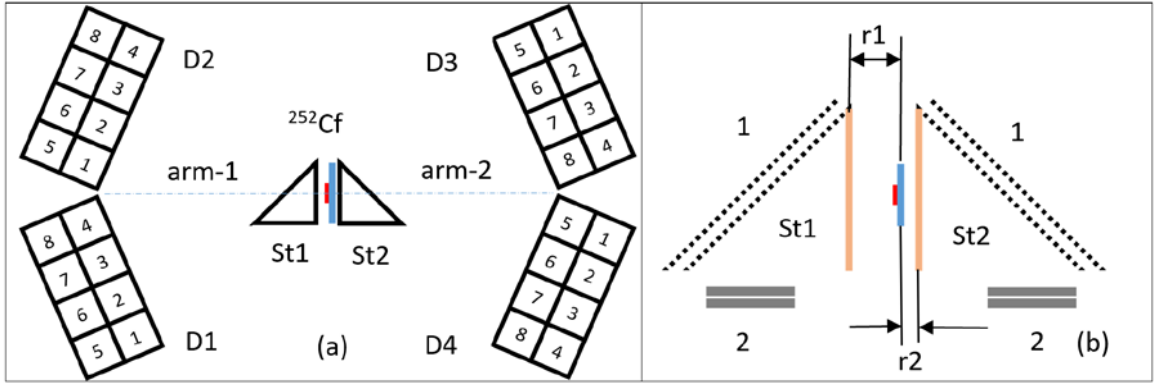

**Fig. 1.** (Color online). Ex1: layout of the experimental setup (a). The $^{252}$Cf (sf) source is placed between two "start" timing detectors on the microchannel plates $St$1 and $St$2. Arm-2 of the spectrometer faces the source backing. Four mosaics ($D$1 ÷ $D$4) of eight PIN diodes each provide measurement of both the FF energy and time-of-flight. The mean flight-path for each mosaic does not exceed 15 cm and the active area of the PIN diode is 1.4×1.4 cm$^2$. Adjacent PIN diodes in the mosaic are separated from each other by a 5 mm thick strip, opaque for fragments. (b) Schematic configuration of source and time pick-off detectors placement. An active spot of the Cf source is shown in red, its backing is shown in blue. The conversion foils are shown in brown. The electrostatic mirrors and microchannel plates are marked by numbers 1 and, 2 respectively. The distances marked by the arrows are r1 = 3 mm, r2 = 0.5 mm.

Calculation of the FF mass with accounting for pulse-height-defect (PHD) was performed in iterative procedure based on parametrization proposed in ref. [17]. The calculation algorithm was presented in ref. [18].

Below the results of Ex1 will be compared with those obtained earlier in Ex2. Configuration of the source and placement of time pick-off detectors were similar to those shown in fig. 1(b). 22 µg/cm$^2$ thick conversions Milar foils with a 40 µg/cm$^2$ thick gold layer on it was used. Corresponding distance r1 does not exceed 6mm. This experiment was described in our previous publications [2, 4].

The setup used in Ex1 had larger aperture (four mosaics of PIN diodes instead of two), different electronics and data processing procedures compared to Ex2. Flash-ADC and CAMAC modules provided data acquisition in Ex1 and Ex2, respectively. The FF mass reconstruction procedure in Ex1 was based on a direct accounting for both PD and PHD while interpolation of their influence in the energy range from alpha-particles up to FFs was utilized in Ex2 [2].

## 3 RESULTS

FFs mass correlation distribution in the region of the "Ni-bump" [2] obtained in Ex1 is presented in fig. 2(a). The masses of the fragments detected in arm-1 and arm-2 are marked as $M_1$ and $M_2$ respectively. Due to the actual background conditions, *the events with the energy of the light fragment in the range $E_2 = (6 \div 30)$ MeV were selected*. The reason for such choice is illustrated in (fig. 4(a)) below. Similar distribution from Ex2 is shown in fig. 2(b). The projections of the mass correlation distributions onto $M_2$ axis are compared in fig. 2(c). Fig. 2(a) shows clearly the vertical "bridge" that "connects" the horizontal lines $M_2 \approx 68$ u and $M_2 \approx 72$ u at $M_1 \approx 144$ u. Some points in the bottom left corner of the Ni-bump lying between the same lines are associated with the tilted line $M_1 + M_2 =$ const. These nine points were omitted when creating fig. 2(c) (curve in black). As can be inferred from fig. 2(c), the statistics in Ex1 are approximately three times higher than that of Ex2. A total yield of the events in the box $M_1 = (120 \div 156)$ u, $M_2 = (55 \div 75)$ u in Ex1 does not exceed $1.8 \times 10^{-4}$ per binary fission which is similar to previously obtained value.

As can be inferred from fig. 2(c), the peak centered at mass 72 u has experimental *fwhm* ≈ *1.5 u*. The largest contribution to this value is given by the relative timing resolution, which is acceptable despite the short flight-path. Due to the energy selection mentioned above, the velocities of the Ni fragments lie in the vicinity of *0.4 cm/ns*. This velocity is *three times less* than the velocity of the typical FF of the light mass peak. We observe *completely resolved peaks* for isotopes of $^{69}$Ni and $^{72}$Ni in fig. 2(c). It means that *their centers are ≈ 6σ apart, where* σ = 0.5 u is a standard deviation of the mass peak with *fwhm = 1.2 u*. In order to agree two estimates, less rough partition (0.25 u per bin) should be applied but it is not suitable for the data of Ex2 with lower statistics.

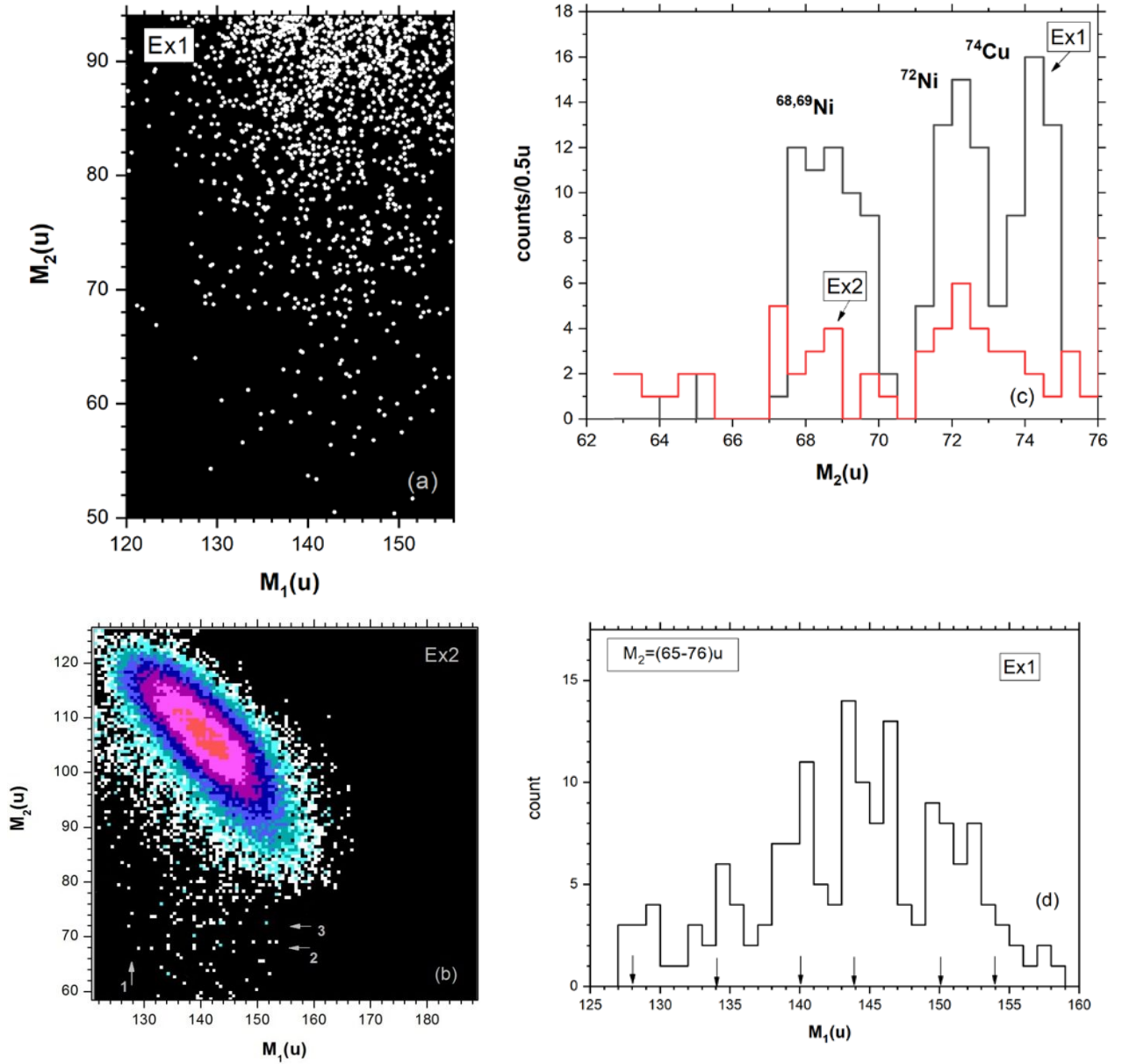

**Fig. 2.** (Color online). FFs mass correlation distributions from $^{252}$Cf(sf) obtained in Ex1 – (a) and Ex2 – (b). The events with the energy of the light fragment in the range $E_2 = (6 \div 30)$ MeV were selected in fig. 2(a). Comparison of their projections onto $M_2$ axis – (c). The total yield of the events in the spectrum in black ≈ $1.3 \times 10^{-4}$ per binary fission. Projection of the distribution obtained in Ex1 onto $M_1$ axis under condition that $M_2 = (65 – 76)$ u – (d). Positions of the magic nuclei are marked by the arrows. It should be noted that (b) was published in refs. [2, 4]. See text for the details.

The above mass resolution derived *directly* from the measured data does not contradict the values of the typical fission fragments (FFs) energy and time resolutions known from the literature. Using the time resolution of the PIN diode reported in ref. [19] at ≈ 0.3 ns, one can calculate that the relative measurement error of the velocity does not exceed 0.8%. The maximal fluctuation of the flight-path (0.3%) is not significant. Energy resolution is roughly estimated as 0.1xPHD (Pulse Height Defect) [20] and does not exceed 0.75% for the PHD ≈ 1.2 MeV, known from the experiment. The resultant mass resolution is estimated to be 1.2 u. We use old literature data for the "pessimistic" estimate. The modern value of the PIN diode time resolution, has the maximum input into mass resolution, is almost order of magnitude better ~ 0.054 ns (fwhm) [21].

The data of Ex2 indicated that the heavy clusters in the ternary pre-scission configurations are predominantly magic nuclei as it was supposed in ref. [4] (see table 1 in work [4]). Noticeably large statistics in Ex1 allowed confirming this assumption. The projection of the distribution shown in fig. 2(a) onto $M_1$ axis for the range of $M_2 = (65–76)$ u (fig. 2(d)) vividly demonstrates increased yield of the heavy fragments, corresponding to the magic isotopes of $^{128}$Sn, $^{134}$Te, $^{140}$Xe, $^{144}$Ba, $^{150}$Ce, $^{154}$Nd (their masses are marked in fig. 2(d) by the arrows).

The nuclear charge of the fragment of the fixed mass is estimated in the frame of the $Z_{UCD}$ (Unchanged Charge Density) hypothesis [22, 23] i.e. it supposed that the ratio $Z/N$ (number of protons/number of neutrons) is the same in the FF and in the mother system.

The data from Ex1 along with the presence of the lines at the mass numbers $A = 128, 68, 72$ (fig. 2(a)) similar to those observed in Ex2 (marked by the numbers 1, 2, 3, respectively in fig. 2(b)), show some additional structure. It consists of the family of lines $M_1 + M_2 \approx$ const at an angle of $45^0$ to *x*-axis (partially presented in Ex2 as well) and some lines almost perpendicular to them. The lines $M_1 + M_2 \approx$ const involved in the rhombic-like structure are discussed in ref. [24].

We have paid special attention to the additional peak centered at the mass $M_2 = 74$ u in Ex1 data (fig. 2(c)). The origin of the peak becomes clear from the analysis of the mass correlation distribution obtained in Ex2 (fig. 3), with less statistics compared to fig. 2(b). Fig. 3 demonstrates the "fine structure" of the "Ni-bump". Line $M_2 = 68$ u, associated with isotope of $^{68}$Ni$_{40}$, starts at $M_1 = 128$ u (presumably $^{128}$Sn) and continues to mass $M_1 = 144$ u (magic isotope of $^{144}$Ba). Then the line switches to the mass $M_2 = 69$ u likely corresponds also to isotope of Ni. Thus, magic proton shell $Z = 28$ and neutron subshell $N = 40$ provides the effect. The next structure starts at the point (134, 70) u and then transforms into the line $M_2 = 65$ u. According to the unchanged charge density hypothesis (UCD hypothesis) $N = 40$ corresponds to $Z = 25$ thus the horizontal line is due to the isotope of $^{65}$Mn$_{40}$. One more structure lies above the structures discussed above. The horizontal line starts at the point (128, 74) u and continues to mass $M_1 \approx 140$ u (magic isotope of $^{140}$Xe) and "jumps" to the mass $M_2 = 72$ u. In the frame of the UCD hypothesis, the masses $M_2 = 74, 72$ u are associated with $^{74}$Cu and $^{72}$Ni nuclei respectively.

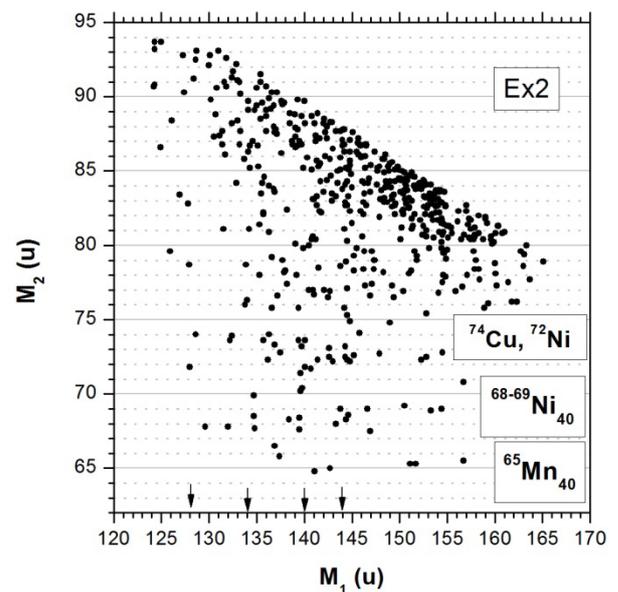

**Fig. 3.** Demonstration of the "fine structure" of the "Ni-bump": mass correlation distribution observed in Ex2 with less statistics compared to fig. 2(b). See text for details.

As was mentioned at the beginning of this section, the events with the energy of the light fragment in the range (6÷30) MeV were selected. It should be added that similar selection was applied in both arms. The reason of such restriction becomes clear from fig. 4(a) that shows the distribution $M_1$–$V_1$ in the arm-1. The loci of conventional binary fission (number 1 in the figure) are prolongated by the "tails" marked by the numbers 2 ÷ 5. The tails 2 and 3 are due to the FF scattering on the frame bounded the active surface of PIN diode. Scattering of the fragments on the mirror wires gives rise to the tails 4, 5. As can be inferred from the figure, the region *inside the box W at least up to the $M_1 \approx 80u$ is almost free from the background of the scattered events.* Similar distribution and selection box could be shown for the arm-2.

The structures observed in arm-2 (figs. 2(a), 2(c)) could be due in principal to the interaction of the FFs sequentially with the source backing and the conversion foil of the detector St2 (fig. 1). In order to exclude the influence of the latter foil, we obtained the difference spectrum by subtracting the spectrum of the light masses measured in arm-1 from that measured in arm-2. *Thus, the resultant spectrum (fig. 4(b)) is a "pure" effect of the $Al_2O_3$ backing.* Comparing the spectra in fig. 2(c) and Fig. 4(b) we conclude that the Lexan foil in arm-2 does not significantly affect the Ni-bump. Thus, we repeated the same differential approach which let us to reveal the Ni-bump for the first time [1, 2].

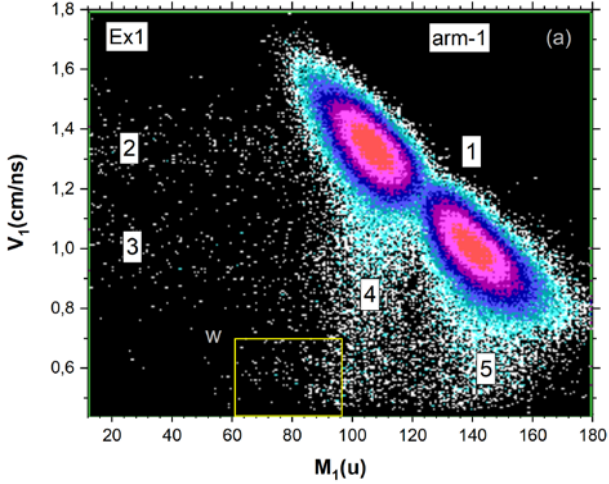
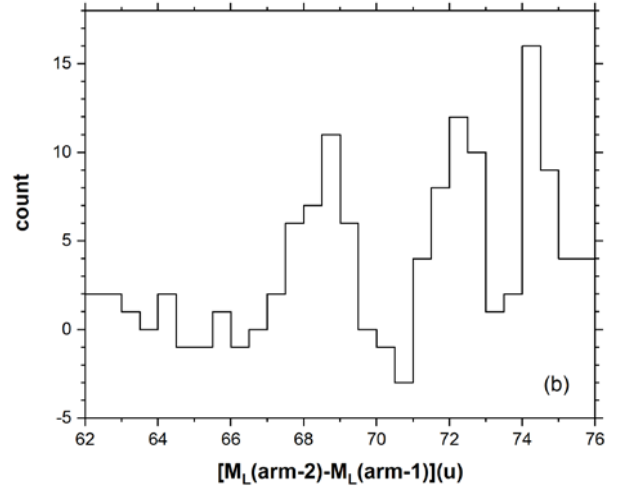

**Fig. 4.** (Color online). The FF mass-velocity distribution in arm-1 (a). The loci of conventional binary fission (number 1 in the figure) are prolongated by the "tails" marked by the numbers 2 ÷ 5. The selection window W is marked by the yellow rectangle. (b) Difference spectrum of the light FF masses measured in Ex1 in arm-1 and arm-2 respectively. See text for details.

## 4 DISCUSSION

### 4.1 Population of the shape isomer state in the intermediate fragment

The analysis of the energies of the FFs linked to Ni-bump allowed us to come to conclusion that the first rupture occurs in a very elongated configuration of the decaying system [4]. The fact of the population of such states in Cf fission is confirmed in a recent publication [25].

Calculations of the potential energy surface of the $^{252}$Cf nucleus [26] performed using the Strutinsky prescription showed that at large deformations of the nucleus there are two distinct potential valleys. In both valleys, after the rupture, at the narrowest section of the neck, almost all the deformation energy concentrates in the heavy or light fragment respectively (figs. 5(a), 5(b)). Such asymmetric partition of the deformation energy is confirmed by the neutron data [27]. Recent calculations of the potential energy surfaces for some super-heavy nuclei [28] applying the same theoretical approach showed similar shapes of the fissioning system near scission point. The result for the $^{280}$Ds nucleus is presented in fig. 5(c).

Similar to the ternary mass split $^{72}$Ni + $^{52}$Ca + $^{128}$Sn at the left boundary of the Ni-bump (figs. 2(a), 2(b)), the CCT fragments will be dubbed "Ni-like", "Ca-like" and "Sn-like". The light fragment formed after the first rupture (*intermediate fragment*) will be dubbed "Cd-like" nucleus.

The experimental data on the kinetic energies of the CCT partners indicate the sequential or semi sequential character of the process [4]. It means that there is a delay between the consecutive ruptures, at least comparable with the time of full acceleration of the Cd-like fragment, and at the same time this fragment is excited enough to undergo fission. Both requirements could be met if the Cd-like nucleus is formed *in the fission isomer state*. This hypothesis mentioned also in ref. [5, 29] will be analyzed below.

The deexcitation of the shape isomer states in some FFs by radiation of x-ray quanta was reported in ref. [30]. Multiple local minima at the potential energy surface for the wide range of the heavy nuclei including FFs were found in ref. [31]. These wells could give rise to the shape isomer states.

In both [4] and this work, the excitation energy $E_3^* \approx 30$ MeV of the FFs involved in the "Ni-bump" was estimated according to the formula:

$$E_3^* = Q_3 - \text{TKE}_3 \quad (1)$$

where $Q_3$ is a reaction heat of tri-partition, $\text{TKE}_3$ is a total kinetic energy of three fragments. Bearing in mind that Ni-like and Sn-like fragments are the magic nuclei (fig. 2(b)), almost all $E_3^*$ should be concentrated in the Ca-like fragment. At the same time, very elongated Cd-like fragment (fig. 5(b)) formed after first scission could evolve to dumbbell-like shape that consists of the magic Ni-like and elongated Ca-like parts connected via well-defined neck, and $E_2^*$ is mostly concentrated in the Ca-like nascent fragment (pre-fragment).

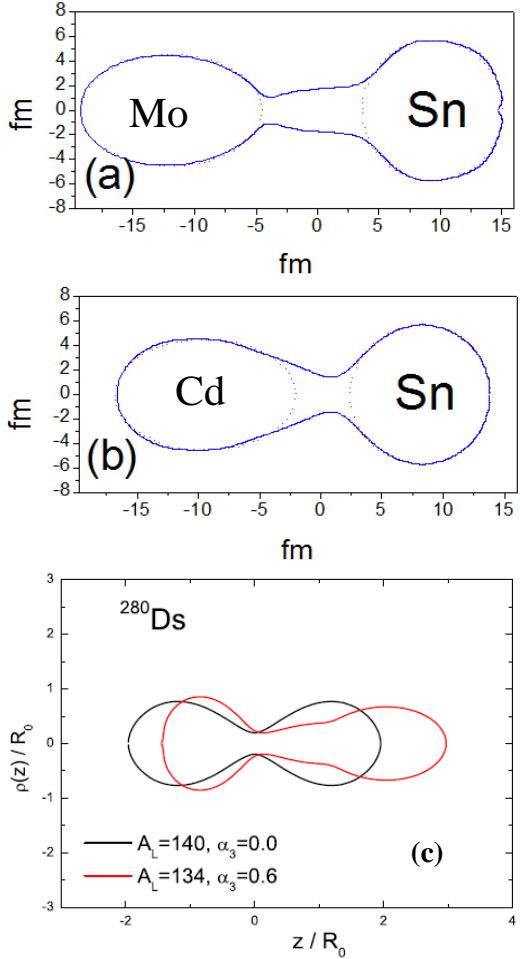

**Fig. 5.** The shapes of the $^{252}$Cf nucleus for large elongations in two different potential energy valleys (a) and (b) [26]. The shapes of the super-heavy $^{280}$Ds nucleus near the scission point for two values of the octupole deformation $\alpha_3$ [28].

Similar shape is predicted, for instance, for the $^{232}$Th nucleus at the third minimum of the fission barrier [32], fig. 6. For the Cd nucleus discussed above, magic Ni and Ca nuclei could play the same role as magic Sn and Zr substructures in $^{232}$Th.

Preformation of the nascent fragments at the top of the fission barrier, at least for the actinides, has been also confirmed in other theoretical approaches [33, 34].

Thus, it is reasonable to assume that $E_3^* \approx E_2^*$ because both values define the deformation energy of the same Ca-like fragment, or pre-fragment, in the ternary and binary fission, respectively.

The value of $E_2^*$ is very close to the fission barrier $E_b$ of the Cd-like nuclei. For instance, for the isotope of $^{109}$Cd, the experimental value of $E_b$ does not exceed 34 MeV (table E in ref. [35]). Experimental data from [36, 37] show that the barriers peak at symmetry and decrease towards both larger and smaller values of the mass asymmetry. For the CCT events under discussion, the mass asymmetry for the different partitions of the Ni/(Ca-like) fragments varies between 0.66 and 0.36, and this corresponds to up to 10 MeV decrease in the fission barrier, for instance, for isotopes of Mo [37]. Therefore, population of the fission isomer state in the Cd-like fragment is not prohibited energetically.

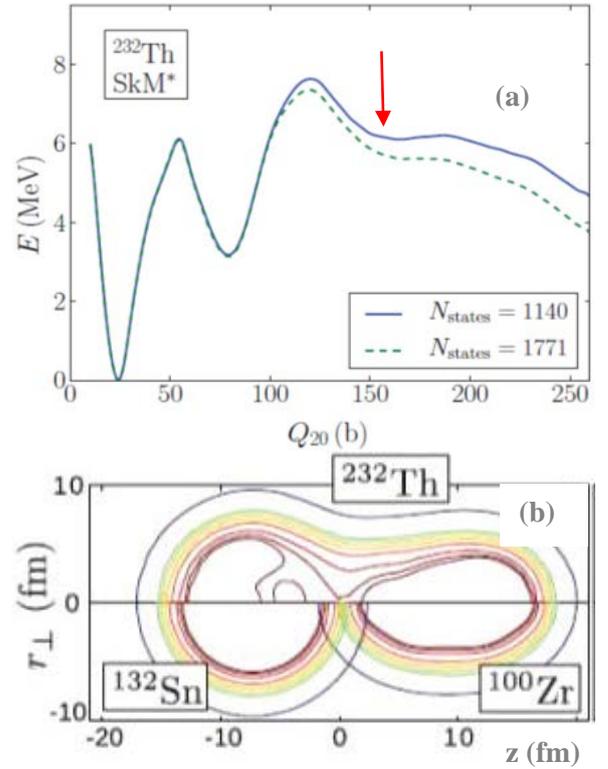

**Fig. 6.** (Color online) Potential-energy curve (fission barrier) for $^{232}$Th calculated in the frame of the finite-temperature superfluid nuclear density functional theory [27] for two different numbers $N$ of the basis states (a). Cross section of total density of $^{232}$Th at third minimum ($Q_{20} = 165$ b) (b).

### 4.2 Correction of the experimental yield of the CCT events

In this work, we try to present maximum detailed experimental information with the aim to attract the attention of the theoretical community for the development of the quantitative model of the CCT process. In this regard it is necessary to refine the "true" yield of the events in the Ni-bump $Y$(Ni) compared to all fission events detected.

The experimental value of the yield $Y_{exp}$(Ni) ≈ 1.3×10$^{-4}$ per binary fission (fig. 2(c)). should be regarded as a lower limit of the yield. Peculiarities of the detection of almost collinear fragments with PIN diodes were examined in ref. [2]. In order to estimate the registration efficiency of detecting of the CCT partners flying in the same direction with the fixed opening angle between them special Monte-Karlo simulation was performed. The results are presented in fig. 7

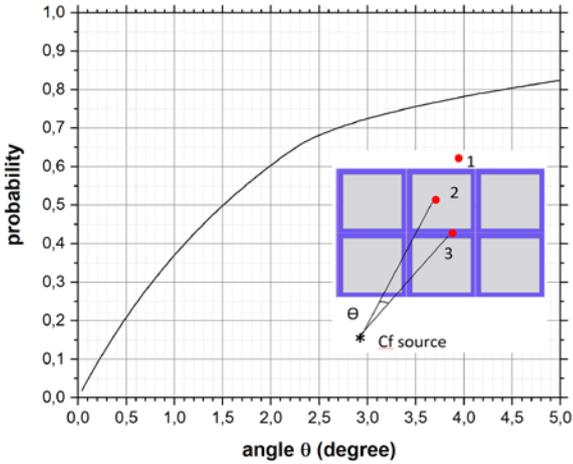

**Fig. 7.** (Color online). Calculated probability of the detection of the CCT event (CCT event/binary fission) in the frame of the missing mass approach. Several PIN diodes from the mosaic with the three labeled different possible locations (red points) a particle can hit the array are shown in the inset. See text for details.

Several PIN diodes from the mosaic with the three labeled different possible locations (red points) a particle can hit the array are shown in the inset. Position 1 is a particle that misses the array completely, position 2 is a particle that hits one of the active surfaces (1.4×1.4 cm$^2$) of the detector array, and position 3 is a particle that hits the border area between the active surfaces. Particle locations 1 and 3 both count as a "miss" and therefore get assigned the score value of 0. Particle location 2 counts as a "hit" and get the "score" value of 1 assigned to it. When simulating two CCT partners with the fixed opening angle θ between them one ends up with 4 possible combinations: (1,1); (1,0); (0,1) (0,0). Here the "score" value of the first particle is the first digit and the "score" of the second particle is the second digit. For example, (1,0) means the 1$^{st}$ particle was in position 2, and the 2$^{nd}$ particle was in either the 1$^{st}$ or 3$^{rd}$ position. Finally define #(x, y) as the total number of events with outcome (x, y). With these terms defined the following relationship describes the relative probability $P$ of the detection of the CCT event (CCT event/binary fission) in the frame of the missing mass approach:

$$P = \frac{\#(1,0) + \#(0,1)}{\#(1,0) + \#(0,1) + \#(1,1)}$$

If Ni- and Ca-like fragments are produced perfectly collinearly after the break-up of the Cd-like fragment, they are dispersed in the source backing in a fork-like manner with an opening angle $\theta_{min}$ ≈ 0.3$^0$. Another limit $\theta_{max}$ ≈ 2$^0$ is determined by the condition of the detection of the fork in the neighboring PIN diodes. The angle θ defines "the active area" along the PIN diode boundaries where one partner of the fork will be detected while the other one will be lost in the detector frame. Given this range of angles and actual dimensions of the setup (fig. 7), $Y$(Ni) = 2.2×10$^{-4}$÷10$^{-3}$. Thus, in any self-consistent theoretical model of the CCT the predicted open angle θ between the CCT partners flying in the same direction and their yield must be correlated within the dependence shown in fig. 7.

### 4.3 Alignment and break-up of a highly deformed fission fragment during its passing through a solid-state medium.

The trivial role of the medium in the way of intermediate fragment consists in deceleration of the fragment which energy losses exceeds some MeV. There exist, however, several more subtle effects.

Some of these effects could result in the alignment of the dipole moment of the fragment along its linear momentum direction. As an example, let us consider the form of an ion decelerated in a medium. The Coulomb interaction of the external electron shells of the ion with the medium is known to be the dominant factor of the energy losses at deceleration of the ion with the energy approximately 1 MeV/A typical for the fission fragments. At rest or when moving in vacuum a nucleus is located in the charge center of the electron shell of an atom or an ion. At deceleration, due to the inertia, the nucleus turns out to be shifted forward from the charge center of the electron shell. Thus, the nucleus is brought into a nonzero electric field directed anti parallel to the decelerating force.

As demonstrated above, the Cd-like intermediate nucleus in some cases is generated in fission process in the shape isomer state. This state looks like a di-nuclear system of Ni and Ca clusters. The shell effects make energetically preferable a certain difference in $Z/A$-ratios of the members of this pair. In this case the dipole moment of such a system appears. This dipole moment interacts with the nonzero electric field of the electron shell. This effect may align the axis of the di-nuclear system to be parallel to linear momentum of the ion. For some part of the Cd-like fragments interacting with the nuclei of the medium at relatively small impact parameters the deceleration-induced dipole moment and, accordingly, the effect under discussion sharply increases.

Other mechanisms of the in-medium alignment should also be subjects of such an analysis. In particular, we have in mind interaction of the large quadrupole moment of the discussed fragment with quadrupole moments of the decelerating medium atoms.

All these mechanisms may be treated by analogy with stabilization mechanisms typical for ballistic problems. Any deviation of the symmetry axis of the prolate body from the direction of a decelerating force causes the restoring force to occur [38].

As was demonstrated in ref. [4], the experimental data support the sequential scenario of the ruptures of the mother nucleus that causes its ternary division i.e. there is a delay between two ruptures. The isomer (shape isomer) state of the intermediate Cd-like fragment is existing during this delay. At the same time, the decay of the Cd-like nucleus cannot be spontaneous. Any in-flight decay would give incorrect measured velocities while we observe well-defined structures and not random seeding of the mass-mass plot (fig. 2(a)). Moreover, the products of such a spontaneous decay would be observed in the form of symmetric bumps in both arms of the experimental setup. Thus, it is reasonable to suppose that the break-up of the Cd-like fragment takes place in the source backing, and so the break-up is an induced decay. It cannot be fission that follows the fusion of the Cd-like fragment with the nucleus from the backing because the interaction energy is much lower than the Coulomb barrier.

The idea of fission induced by Coulomb interaction of the fissioning nucleus with a heavy charged projectile was introduced more than fifty years ago [39]. Further improvement of the theoretical description of the Coulomb fission was performed in refs. [40, 41]. Differential cross-section of about 50 mb/sr for the Coulomb fission of $^{238}$U due to interaction with $^{148}$Nd projectile at energy $E = 500$ MeV in the c. m. frame was predicted in ref. [40]. In our case the masses, charges, shapes and energies of the interacting nuclei are substantially different. In addition, what is particularly important, the state of the Cd-like nucleus formed by a Cf fission event is supposed to be in the second or the third well (in the case that the latter exists), as it has just been demonstrated (see for instance fig. 6). The external barrier of such well is expected to be incomparably lower than the fission barriers of actinides in their ground states. Excited states of such a type could be referred to as shape isomer ones and according to [42] may be treated as di-nuclear systems. These states may be also considered as hyper-deformed ones [43–45], with the only difference that their angular momenta are not extremely large. The masses from the range $A = 105–130$ are considered to be promising to search for hyper-deformed states [44]. Unfortunately, to date, the explicit form of the fission barriers of these nuclei is unknown. One more property of the Cd-like fragment is that due to its great elongation this nuclide possesses very small rotational quantum energy ($\hbar\omega \approx 1–2$ keV). Thus, the intensity of multi-step excitation process of such objects during passing through a source backing medium is high and thereby the probability to overcome the lower external barrier is expected to be high.

Mechanism of the Coulomb fission could be decisive for the alignment under discussion as well. Coulomb excitation is known [46] to result in a strong alignment of the symmetry axis of the fissioning nucleus perpendicular to the line connecting the center of mass of the target and projectile nuclei at the point of the closest approach where fission takes place. In the case under consideration, when the scattering angle of the Cd-like fragment is relatively small, similar orientation effect would provide an alignment of the fragment and its linear momentum at the moment of scission.

It should be noted that all the considerations presented in this section are qualitative ones and therefore need further quantitative verification.

# 5 SUMMARY

The new experimental data with three times larger statistics than before confirms the structure of the "Ni-bump" observed earlier in the spectrometer arm-2 facing the Al$_2$O$_3$ $^{252}$Cf source backing. Projection of the bump onto the axis of the light FFs masses shows the peaks centered at the mass numbers 68, 69, 72, associated with the magic isotopes of $^{68,69,\,72}$Ni. The projection onto the opposite axis of the heavy masses also demonstrates the peaks at the masses, which could be assigned to the magic isotopes of $^{128}$Sn, $^{134}$Te, $^{140}$Xe, $^{144}$Ba, $^{150}$Ce, $^{154}$Nd [47].

For the first time, a specific rhombic-like structure consisting of the lines corresponding to the fixed missing mass and the lines approximately perpendicular to them was revealed in the FFs mass correlation distribution below the Ni-bump.

The following scenario gives a self-consistent interpretation of the Ni-bump features observed. In this mode, the CCT occurs as a two-stage sequential decay process of the very deformed pre-scission configuration (fig. 5(b)). The first rupture forms the shape isomer state of Cd-like nucleus in the second or third well near the top of its fission barrier. Thus, similar to the known fission isomers in actinides, the Cd nucleus in this state could be dubbed as *fission isomer of Cd*. It should be stressed that we observe *induced fission of Cd-like nucleus from the weakly-bound shape isomer state*. This nucleus (intermediate fragment) undergoes a Coulomb break-up while crossing the Cf source backing.

The interaction of the intermediate fragment with the source backing is appeared to be also the origin of collinearity of the CCT partners.

**Acknowledgments** This work was supported, in part, by the Russian Science Foundation and fulfilled in the framework of MEPhI Academic Excellence Project (Contract No. 02.a03.21.0005, 27.08.2013), and by the Department of Science and Technology of the Republic of South Africa (RSA). We are grateful to R.V. Jolos, N.V. Antonenko, G.G. Adamian, V.I. Furman, A.K. Nasirov, N. Carjan for their continuous interest in the subject and the stimulating discussions.